\documentclass[12pt]{article}

\usepackage{amsfonts}
\usepackage{amsmath}
\usepackage{amssymb}
\usepackage{graphicx}
\usepackage{epsfig}

\begin{document}
%\preprint{LEZ/00104}

\title{Formation of ions by high energy photons}

\author{E. G. Drukarev $^{a,b}$ , A. I. Mikhailov$^{a,b}$, I. A. Mikhailov$^{a,b}$, \\
Kh. Yu. Rakhimov$^{c,b}$, and W. Scheid$^b$ \\
$^a$ \em Petersburg Nuclear Physics Institute,\\ \em Gatchina, St. Petersburg 188300,
Russia\\
$^b$
\em Justus-Liebig-Universit\"at Giessen, Giessen 35392, Germany \\
$^c$ \em Heat Physics Department of the Uzbek Academy of Sciences, \\ \em Tashkent 700135, Uzbekistan}

\date{October 26, 2006}
\maketitle

%\keywords{one two three}

\begin{abstract}

We calculate the electron energy spectrum of ionization by a high energy photon,
accompanied by creation of $e^-e^+$ pair. The total cross section of
the process is also obtained. The asymptotics of the cross section does not depend on the
photon energy. At the photon energies exceeding a certain value $\omega_0$ this
appeares to be the dominant mechanism of formation of the ions. The dependence of
$\omega_0$ on the value of nuclear charge is obtained. Our results are consistent with
experimental data.

\end{abstract}

\section{Introduction}

In the present paper we calculate the cross sections $\sigma$ for formation of
ions in interactions of high energy photons with atoms. We calculate also
the distribution $d\sigma/d\varepsilon$ for the process in which the final state contains
an ion and electron with energy $\varepsilon$. We consider the high energy asymptotics
of these characteristics, i.e. we consider the photon energies
\begin{equation}
\omega \gg m,
\end{equation}
with $m$ standing for the electron mass at rest(we employ the system of units with $\hbar=c=1$).
We shall include only the lowest terms of expansion in $\omega^{-1}$ in our calculations.

The simplest mechanism for formation of ions is the photoionization (photoeffect),
in which the final state
consists of an ion and a continuum electron. It is also known that while the energies
increase, the Compton scattering on the bound electron dominates. Ionescu {\em et al}
\cite{1} noted that at still higher energies ions are produced mainly being accompanied by
creation of electron-positron pairs
and provided estimates for the corresponding cross section.

Here we carry out the calculations for the distributions $d\sigma/d\varepsilon$ and for the cross section
$\sigma$. We focus on the case of not very large values of nuclear charge
\begin{equation}
 (\alpha Z)^2 \ll 1,
\end{equation}
adding, however, an analysis of the case, when $(\alpha Z)^2$ is not considered as a small parameter.
When unequality (2) is true, we can separate three scales of the electron kinetic energies $\varepsilon$.
Besides the characteristic values of $\omega$ and $m$, the third one is the electron binding energy $I_b$.
For K-shell in the hydrogenlike approximation $I_b=I=\eta^2/2m$, with $\eta=m\alpha Z$.

We demonstrate  that for the energies $\varepsilon \sim m $ the distribution $d\sigma/d\varepsilon$ is determined by the
vacuum assisted mechanism. The ionized electrons can be distinguished from those of $e^-e^+$ pair since the latter carry mostly energies
$\varepsilon_i \sim \omega \gg m$. We show how the distribution can be presented in terms of the pair creation on a free electron at rest.
We show also that at these values of the electron energies the distribution
does not depend on the details of atomic structure.
At $\varepsilon \ll m$ the distribution behaves as $\varepsilon^{-1}$. This means that in order to calculate
the cross section $\sigma$ one has to include the region $\varepsilon \sim I_b$ which should be treated separately.
We show that in the asymptotics (1) the cross section reaches a constant value and calculate it.

%We show that ionization,
%preceded by the pair production in the field of the nucleus
%also discussed earlier in \cite{1}
% provides the leading correction of the order $(\alpha Z)^2\ln(\omega/m)$
%the distributions obtained in the paper. This is also the main mechanism for knocking
%an electron with the energy $\varepsilon_2 \gg m$ from the atom.

In Sec.2 we present general equations. In Sec.3 we calculate  distribution $d\sigma/d\varepsilon$ for $\varepsilon \gg I_b$.
In Sec.4 we calculate this distribution for $\varepsilon \sim I_b $.
In Sec.5 we carry out a matching of the energy distributions in the two regions.
In Sec.6 we calculate the cross section of ion production
$\sigma$ and compare the results of our calculations with experimental data.
We summarize in Sec.7.

\section{Notations and general formulas}

In all the processes considered in the paper an electron is removed from the atom to continuum.
It will be instructive to consider simultaneously a similar process on the free electron at rest.
For the latter case we denote the four-momenta of the electrons as $p_{1,2}$ with the time component
$p_{10}=m$ and $\bf p_1=0$. We denote four-momenta of the electron and positron
of the created $e^-e^+$ pair as $p_e$ and $p_p$ correspondingly.
For each $p_i(i=e,p,1,2)$ the total energy is $E_i=\sqrt{m^2+\bf p_i^2}$, while
kinetic energies are $\varepsilon_i=E_i-m$.
The four-momentum of the incoming photon is $k$, while for its three-dimension momentum
we can write $|\bf k|=\omega$.

In the pair production on the nucleus the latter accepts linear momentum
\begin{equation}
 \bf q=\bf k- \bf p_e-\bf p_p.
\end{equation}
In the pair production on free electron momentum (3) is transferred to the latter. In pair production on the
bound electron momentum
\begin{equation}
 \bf Q=\bf q-\bf p_2,
\end{equation}
is transferred to the residual ion.

The cross section for pair creation in the field of the nucleus was first calculated by Bethe and Heitler
$\cite{2}$. It
can be represented as
\begin{equation}
 d\sigma_{BH}=\frac{\pi}{\omega}|F_{BH}(\omega, p_e,p_p)|^2\delta(\omega-E_e-E_p)d\Gamma,
\end{equation}
with the phase volume
$d\Gamma=\frac{d^3p_{e}}{2E_e(2\pi)^3}\frac{d^3p_{p}}{2E_p(2\pi)^3}$.

At $\omega \gg m$ it is convenient to present momenta $p_i$ ($i=e,p$) as $p_i=(E_i, p_{iz}, \mathbf p_{it})$
with the axis $z$ directed along photon momentum $\bf k$. The lower index $t$ denotes the components,
which are orthogonal  to $\bf k$.
The energy distributions are determined by small $p_{it} \sim m$ ($i=e,p$) \cite{2}, \cite{3}
with $q$ being determined by Eq.(3).
Thus presenting
\begin{equation}
E_i=|p_{iz}|+\frac{m^2+p_{it}^2}{2|p_{iz}|},
\end{equation}
we find that
the longitudinal component of the recoil momentum $\bf q$ is $m/\omega$ times smaller than the transverse one.
Hence we can write
\begin{equation}
 \bf q=- \bf p_{et}-\bf p_{pt}
\end{equation}

For the pair creation in the field of the bound electron the cross section is
\begin{equation}
 d\sigma=\frac{\pi}{\omega}|F(\omega, p_e,p_p, p_2)|^2\delta(\omega-E_e-E_p-E_2+m-I_b)d\Gamma_{b},
\end{equation}
with $F$ being the amplitude of the process, while
$d\Gamma_{b}=d\Gamma\frac{d^3p_2}{2E_2(2\pi)^3}$,
$I_b$ stands for the ionization potential of the bound state.

To avoid writing complicated expressions which describe the third order amplitudes $F_{BH}$ and $F$ we
present them by the Feynman diagrams, following \cite{1}.
The pair creation in the field of the nucleus is shown by the diagrams of Fig.~1.
The pair creation on the bound electron is shown by the diagrams of Fig.~2.
The possible permutations of the final state electrons should be added.
The diagram $a$ of Fig.~2 shows creation of pairs by the photon with further scattering on the bound electron. In the
diagram $b$ of Fig.~2 the photon is initially absorbed by the bound electron with further radiation of a photon which
creates the electron-positron pair.

\section{Fast ionized electrons}.

Here we consider the case of fast ionized electrons with the energies
\begin{equation}
\varepsilon_2 \gg I,
\end{equation}
with $I$ being ionization
potential of the K shell electron. We focus on the energies $\varepsilon_2 \ll \omega$,
since the these values provide the leading contribution to the cross section.

The electrons with the energies $\varepsilon_2 \ll \omega$ can come from $e^-e^+$ pairs and
also can be caused by removing of the bound electrons to continuum. In the former case the
distribution $d\sigma/d\varepsilon$ drops as  $\omega^{-1}$ \cite{2,3}. We shall see that
in the latter case it
does not depend on $\omega$. Thus the electrons with energies $\varepsilon_2 \ll \omega$
are mainly those which are knocked out from the atom.

Momentum $Q$ should be transferred to residual ion. It can be transferred by the
initial state bound electron or by final state continuum lepton (electron or positron). It is known \cite{3} that
each interaction of the continuum electron with residual ion provides a factor $\alpha Z E_i/p_i$. On the
other hand, in the bound state wave function momentum $Q$ is compared to the bound state characteristic
momentum $\eta_b$. The wave function reaches its largest
values at $Q \sim \eta_b$, being strongly quenched at $Q \gg \eta_b$. Thus in the amplitude $F(p_i,Q)$
(with $p_i$ denoting momenta of the outgoing electrons and that of positron)
we can neglect $Q$ everywhere except the bound state wave function. This enables to tie the amplitude of the
process on the bound electron with that on the free electron $F_0(p_i)$, known as triplet production.
Such interpretation of the processes
on the bound electrons reflects the ideas of Bethe \cite{5} (see also analysis presented in \cite{6}).

Assuming that the bound electron is described by a
single-particle wave function $\psi(r)$ we can write for the amplitude of the pair creation on the bound electron
\begin{equation}
F(p_i,Q)=\psi_F(Q)F_0(p_i).
\end{equation}

Here
\begin{equation}
\psi_F(Q)=\int d^3r\psi(r)\exp{(-i(\bf Q \bf r))},
\end{equation}
is the Fourier transform of the wave function $\psi(r)$ also referred to as the wave function in momentum space.
To simplify notations we shall omit the lower index $F$ further.

Note that the outgoing electrons can be described by plane waves due to small value $Q \ll m$ of momentum transferred to
the nucleus. In other words, this is due to existence of kinematical region, where similar process on a free
electron could take place. For example, there is no such kinematical region for the photoeffect, and the plane
wave description for relativistic energies is not sufficient \cite{3}.
See more recent discussion in \cite{6}.

Replacing $d^3p_e$ by $d^3Q$ in the phase volume
 we find for the cross section
\begin{eqnarray}
\nonumber
 d\sigma=\frac{\pi}{\omega}|F_0(\omega, p_e,p_p, p_2)|^2
\delta(\omega-E_e-E_p-E_2+m-I_b) \cdot \\
\frac{1}{2E_e}\frac{d^3p_p}{2E_p(2\pi)^3}\frac{d^3p_2}{2E_2(2\pi)^3} |\psi(Q)|^2\frac{d^3Q}{(2\pi)^3},
\end{eqnarray}
Neglecting $I_b$ with respect to $m$ and $\omega$ we find
\begin{equation}
 d\sigma=d\sigma_0|\psi(Q)|^2\frac{d^3Q}{(2\pi)^3},
\end{equation}
with $\sigma_0$ being the cross section of the process on free electron. After integration over $Q$
providing $\int|\psi(Q)|^2\frac{d^3Q}{(2\pi)^3}=1$ (normalization condition) we find the distributions $d\sigma$ to be equal to those
of the process on the free electron.

The triplet production was much studied
\cite{7,8,M,mo,H}.
%\cite{7}, \cite{8},\cite {M}, \cite{mo}, \cite{H}.
It was shown in \cite {7}, that in the considered region the diagrams of
Fig.2b as well as the exchange diagrams of Fig.2a can be neglected.
Analytical expression for the differential distribution integrated over the variables of the
$e^-e^+$ pair was obtained in \cite{8}. The leading term of expansion in powers of $\omega^{-1}$ for the distribution of the
ionized electron can be written as
\begin{equation}
 \frac{d\sigma}{d\varepsilon_2d\Delta^2}=n_e\alpha^3W(\varepsilon_2, \Delta^2).
\end{equation}
Here $\Delta^2=(p_e+p_p)^2$, it can be expressed
in terms of variables of the ionized electron as
$\Delta^2=-2\varepsilon_2(\omega+m)+2\omega p_2t_2$; $t_2=(\mathbf k \cdot \mathbf p_2)/\omega p_2$, $n_e$ stands
for the number of the
bound electrons in the atom. Evaluating Eq.(2.4) of \cite{8} we find
\begin{equation}
W(\varepsilon_2, \Delta^2)=\frac{A(\varepsilon_2,\Delta^2)}{\varepsilon_2B(\varepsilon_2,\Delta^2)},
\end{equation}
with
\begin{equation}
B(\varepsilon_2,\Delta^2)=(\Delta^2+2m\varepsilon_2)^2,
\end{equation}
and
\begin{equation}
A(\varepsilon_2,\Delta^2)=4\beta(1-L+\frac{4m(\Delta^2(m-4\varepsilon_2)+L(2m^2(2\varepsilon_2+m)+
\Delta^2(\varepsilon_2-m)))}
{B(\Delta^2,\varepsilon_2)}).
\end{equation}
Here $\beta=[\Delta^2-4m^2)/\Delta^2]^{1/2}$, and $L=\frac{1}{\beta}\ln\frac{1+\beta}{1-\beta}$.

Since the variable $\Delta^2$ can be viewed as the squared energy of $e^-e^+$ pair in their c.m. frame, we can write the
limitation $\Delta^2  \geq 4m^2$. The upper limit is $\Delta^2=2m\omega$ \cite{8}. The energy distribution can be obtained
by integration of the differential cross section (15) over $\Delta^2$ in these limits. The value is determined by the lower
limit of $\Delta^2$, providing
\begin{equation}
 \frac{d\sigma}{d\varepsilon_2}=\frac{n_e\cdot \alpha r_e^2}{m}T_f(\frac{\varepsilon_2}{m}),
\end{equation}
with $r_e=\alpha/m$, and (the lower index $f$ comes from "fast")
\begin{eqnarray}
\nonumber
T_f(x)=\frac{2}{x}[-\frac{x^3+x^2+2x-1}{x^2(2+x)^2}+\frac{2(2x^4+7x^3+16x^2+5x-3)}{3x^{5/2}(2+x)^{5/2}}
\ln(\sqrt {x/2}+\sqrt{ x/2+1})\\
-\frac{2(1-4x)}{15}\ _2F_1(2,4,\frac{7}{2},-\frac{x}{2})],
\end{eqnarray}
Here $x=\varepsilon_2/m$.
The function $T_f(x)$ is shown in Fig.~3. At $x \ll 1$ we find
\begin{equation}
T_f(x)=\frac{14}{9}\cdot\frac{1}{x}.
\end{equation}
Note that Eq.(19) is not true for $\varepsilon_2 \sim \omega $. Using Eq.(6) one can see that momentum transferred to
the residual ion can not be made as small as
$Q \ll m$ if all
the three final state leptons carry the energies $\varepsilon_i \sim \omega$.

Thus Eq (19) is true for $I \ll \varepsilon_2 \ll \omega$. Since for $\varepsilon_2 \ll m$ the distribution behaves as
$\varepsilon_2^{-1}$, the region $\varepsilon_2 \sim I$ provides a contribution of the same order of magnitude
to the total cross section.

\section{Slow ionized electrons}.

Now we consider the case $\varepsilon_2 \sim I_b$.
In this case the outgoing electron carries momentum $p_2$ of the order of the binding
momentum $\eta_b = (2m\varepsilon_2)^{1/2} \ll m$. Thus momentum $Q$ transferred to residual ion can be as small
as $\eta_b$ only if momentum $q$ transferred to the atom is also of the order of $\eta$.
Hence the amplitude of the ionization shown in Fig.~2a can be written as
\begin{equation}
F(p_i,q)= \frac{1}{Z}F_{BH}(p_i,q)\Phi_b(p_2,q),
\end{equation}
with
\begin{equation}
\Phi_b(p_2,q)=  \int d^3r\psi^*_{\bf p_2}(r)\psi_b(r)\exp(-i(\bf q \cdot \bf r)).
%  = \int d^3f\psi^*_{p_2}(f)\psi_b(f-q).
\end{equation}
%Here the second equality is written in momentum space.
Here $\psi_{\bf p_2}$ is the wave function of the outgoing electron with asymptotical momentum ${\bf p_2}$,
$b$ denotes the state of the bound electron.
Since $p_2 \sim \eta_b$, the interactions between the outgoing electron and the residual
ion can not be treated perturbativly.

Thus we can write
\begin{equation}
d\sigma=\frac{1}{Z^2} d\sigma_{BH}|\Phi_b(p_2,q)|^2\frac{d^3p_2}{(2\pi)^3}.
\end{equation}
The Bethe-Heitler distribution
$$ d\sigma_{BH}=R d\Gamma' ; \quad d\Gamma'=dE_pp_{et}dp_{et}p_{pt}dp_{pt}d\varphi,$$
with \cite{2,3}
\begin{equation}
R=\frac{8\alpha r_e^2Z^2E_eE_p}{\pi q^4 \omega^3}H ,
\end{equation}
and
\begin{equation}
H=-\frac{\delta_{-}^2}{(1+\delta_{-}^2)^2}-\frac{\delta_{+}^2}{(1+\delta
_{+}^2)^2}+\frac{\omega^2}{2E_eE_p}
\frac{\delta_{-}^2+\delta_{+}^2}{(1+\delta_{-}^2)(1+\delta_{+}^2)}+
(\frac{E_e}{E_p}+
\frac{E_p}{E_e})\frac{\delta_{-}\delta_{+}\cos\varphi}
{(1+\delta_{-}^2)(1+\delta_{+}^2)},
\end{equation}
(here we denoted $\delta_{-}= p_{et}/m$,  $\delta_{+}= p_{pt}/m$)
should be evaluated for $p_{it} \sim m$, as in Bethe-Heitler case, but now we need also $q \sim \eta_b\ll m $
This means that $|p_{et}-p_{pt}| \sim \eta_b \ll p_{et,pt}$ and $|\pi-\varphi| \sim \eta_b/m \ll 1$.
Introducing variables $q$ and $t=|p_{et}-p_{pt}|/q$ ($0\leq t \leq 1$) instead of $\varphi$ and $p_{et}$
we can present Eq.(25) in the form
\begin{equation}
H=
\frac{q^2}{m^2(1+\delta_{+}^2)^2}
(\Lambda+\frac{4\delta_{+}^2t^2}{(1+\delta_{+}^2)^2}),
\end{equation}
with
\begin{equation}
\Lambda=\frac{E_e^2+E_p^2}{2E_eE_p} \quad (E_e=\omega-E_p),
\end{equation}
while the phase volume in Eq.(23) becomes
\begin{equation}
d\Gamma'=dE_pp^2_{pt}dp_{pt}\frac{dt}{2(1-t^2)^{1/2}}dq^2.
\end{equation}
After integration over the positron variables and over $t$ we find
\begin{equation}
d\sigma=\frac{14}{9}\alpha r_e^2
\sum_b|\Phi_b(p_2,q)|^2\frac{dq^2}{q^2}\frac{d^3p_2}{(2\pi)^3}.
\end{equation}
The factors $\Phi_b(p_2,q)$ turn to zero at $q=0$ due to orthogonality of the wave functions involved.
Thus $\Phi_b(p_2,q)$ contains $q$ as a factor at $q\rightarrow 0$, and the integral over $q$ on the
right hand side (RHS) of Eq.(29) provides a finite value.

The factors $\Phi_b(p_2,q)$ have been computed for many cases in connection with  electron-atom scattering.
Here we present calculations with nonrelativistic Coulomb functions. We shall provide results for K shell
electrons. Thus the further results of this Section are actually true for the ground states of relatively light
(Eq.(2)) single-electron ions.

Straightforward calculations provide
\begin{equation}
|\Phi_K(p_2,q)|^2=2^8 \pi \cdot N^2 \exp(2\xi \gamma)\cdot
\frac{\eta^5[(q^2-({\mathbf p_2} \cdot {\mathbf q} ))^2+\xi^2({\mathbf p_2} \cdot {\mathbf q} )^2]}{a^4|b|^2}.
\end{equation}
Here $\eta=m\alpha Z$ is the averaged momentum of the K electron,
\begin{equation}
\xi=\frac{\eta}{p_2}=\sqrt\frac{I}{\varepsilon_2},
\end{equation}
is the Sommerfeld parameter of interaction between the outgoing electron and the nucleus.
The other notations in Eq.(30) are $a=({\mathbf p_2}-{\mathbf q})^2+\eta^2$, $b=q^2-(p_2+i\eta)^2$ while
$\gamma=\arg b =\arg(q^2+\eta^2-p_2^2-2i\eta p_2)$. The factor
\begin{equation}
N^2=N^2(\pi\xi)=\frac{2\pi \xi}{1-\exp(-2\pi\xi)}
\end{equation}
is the squared
normalization
factor of the outgoing electron wave function.

Presenting the phase volume of the outgoing electron as $d^3p_2/(2\pi)^3=mp_2d\varepsilon_2d\Omega/(2\pi)^3$,
we can carry out integration over the solid angle $\Omega$
\begin{equation}
\int \frac {d\Omega}{(2\pi)^3}|\Phi_K(p_2,q)|^2=q^2X(p_2,q),
\quad X(p_2,q)=
\frac{2^7}{3\pi}\cdot N^2 \exp(2\xi \gamma)\cdot
\frac{u(p_2,q)}{v(p_2,q)}.
\end{equation}
Here $u(p_2,q)=\eta^5(p_2^2+3q^2+\eta^2)$, $v(p_2,q)=[(q^2-p_2^2)^2+2\eta^2(q^2+p_2^2)+\eta^4]^3$.
Combining Eqs.(29) and (33), we can write
\begin{equation}
\frac{d\sigma}{d\varepsilon_2}=\frac{14}{9}\alpha r_e^2mp_2\int dq^2X(p_2,q).
\end{equation}

Introducing
\begin{equation}
\epsilon=\frac{\varepsilon_2}{I}=\xi^{-2},
\end{equation}
 we find
\begin{equation}
\frac{d\sigma}{d\epsilon}=\frac{14}{9}\alpha r_e^2K(\epsilon),
\end{equation}
with
\begin{equation}
K(\epsilon)=\frac{2^6}{3\pi\xi}N^2\int_0^\infty dx
\frac{ e^{-2\gamma_1/\sqrt\epsilon}(\mu+3x)}
{(x^2+2\nu x+\mu^2)^3},
\end{equation}
with $\mu=1+\epsilon$, $\nu=1-\epsilon$, and  $\gamma_1=\arg(x+\nu+2i\sqrt\epsilon)$.

Thus the energy distribution can be presented as
\begin{equation}
\frac{d\sigma}{d\varepsilon_2}=\frac{\alpha r_e^2}{I}T_s(\epsilon),
\end{equation}
(the lower index $s$ comes from "slow") with $I=m(\alpha Z)^2/2$ being the K-electron binding energy, and
\begin{equation}
T_s(\epsilon)=\frac{14}{9}K(\epsilon).
\end{equation}
The function $T_s(\epsilon)$ is shown in Fig.~4.

Since $N^2 \sim \epsilon^{-1/2}$ at $\epsilon \rightarrow 0$ (see Eq.(32)), we find a nonzero value for
\begin{equation}
K(0)=1-\frac{7}{3}\exp(-4)\approx 0.957.
\end{equation}
Using Eq.(32) one can present Eq.(37) also as
\begin{equation}
K(\epsilon)=\frac{2^7}{3(1-e^{-2\pi/\sqrt \epsilon})}J(\epsilon), \quad
J(\epsilon)=\int_0^\infty dx\frac{e^{-2\gamma_1/\sqrt\epsilon}(\mu+3x)}
{(x^2+2\nu x+\mu^2)^3}.
\end{equation}

At $\varepsilon_2 \gg I$ Eq.(31) provides $\xi \ll 1$, and thus $\epsilon \gg\ 1$. The lowest order of expansion
in powers of $\xi$, corresponding to the plane wave description of the outgoing electron leads to
\begin{equation}
K(\epsilon)=\frac{1}{\epsilon},
\end{equation}
Thus for $I \ll \varepsilon_2 \ll m$
\begin{equation}
T_s(\epsilon)=\frac{14}{9}\cdot \frac{1}{\epsilon},
\end{equation}
in agreement with nonrelativistic limit $\varepsilon_2 \ll m$ of Eq.(18) (see Eq.(20)).

We see that the lowest order expansion in powers of $\xi$ of the RHS
of Eq.(37) leads
to the same result as provided by nonrelativistic limit of Eq.(18). On the other hand the function
$K$ depends on $\xi$ in terms of parameters $\epsilon=\xi^{-2}$, containing also explicit dependence on
the parameter $\pi\xi$. The latter thus include
the terms which are linear in $\xi$, containing also numerically large coefficient.
We show, however that in our case dependence on $\pi\xi$ cancels out at least in the lowest order
terms of $\xi^2$ expansion.

On the RHS of Eq.(37) dependence on $\pi\xi$ is contained explicitly in normalization factor $N^2$
determined by Eq.(32).
Such dependence
comes also from the exponential factor of the integrand of $J(\epsilon)$ determined by Eq.(41).
Since
$$ J(\epsilon)=\int_{1-\epsilon}^\infty  dy(4\epsilon+3y-2)\frac{\exp(-2\gamma_1/\sqrt\epsilon)}
{(y^2+4\epsilon)^3},$$
(we denoted $y=x-\epsilon+1$)
is dominated by
$y \approx 2\sqrt\epsilon$
we can replace it  by
\begin{equation}
J_1(\epsilon)=\int_{-\infty}^\infty  dy(4\epsilon+3y-2)\frac{\exp(-2\gamma_1/\sqrt\epsilon)}
{(y^2+4\epsilon)^3},
\end{equation}
making the relative error of the order $\xi^{-5} \ll 1$.  Since
\begin{equation}
\gamma_1=\arctan(\frac{2}{y\xi}) \quad at~ y>0;  \quad
\gamma_1=\pi-\arctan(\frac{2}{|y\xi|}) \quad at~ y<0,
\end{equation}
($\xi=\epsilon^{-1/2}$) while for any $x>0$
\begin{equation}
\arctan x=\frac{\pi}{2}-\arctan x^{-1},
\end{equation}
we can write
\begin{equation}
J_1(\epsilon)=\exp(-\pi\xi)\int_0^\infty  dy(4\epsilon+3y-2)\frac{\exp(2\xi \arctan(y\xi/2))}
{(y^2+4\epsilon)^3}
\end{equation}
Integral (47) can be evaluated analytically
\begin{equation}
J_1(\epsilon)=\frac{3}{2^6}\exp(-\pi\xi)\sinh(\pi\xi)\frac{1}{\epsilon+1}.
\end{equation}
Using Eq.(31) we find that the total dependence of the
energy
distribution on parameter $\pi\xi$ cancels out. The limiting Eq.(42) for $\epsilon \gg 1$ can be written as
\begin{equation}
K(\epsilon)=\frac{1}{\epsilon+1}(1+O(\epsilon^{-5/2})).
\end{equation}
Being more rigorous we should replace $1/(1+\epsilon)$ by $1-\epsilon +\epsilon^2$.

Thus several next to leading order corrections to the high energy limit of the function $T_s(\epsilon)$ can be
included by a simple factor
\begin{equation}
g(\epsilon)=\frac{\epsilon}{\epsilon+1}.
\end{equation}
As one can see from Fig.4b the function
\begin{equation}
\tilde{T_s}(\epsilon)=\frac{14}{9}\cdot \frac{1}{\epsilon+1},
\end{equation}
approximates the function (39) well enough even at $\varepsilon_2$ close to zero. The largest relative
deviation
between the RHS of Eqs.(39) and (51) takes place at $\epsilon_2=0$, being about $4\%$.

\section{Matching of the two regions}

The function $T_f$ determined by Eq.(19) describes the energy distribution at $\varepsilon_2 \gg I$ and does
not include corrections of the order $\epsilon^{-1}$. On the other hand the function $T_s$ (39) describes the energy
distribution at $\varepsilon_2 \ll m$ since the outgoing electron is treated nonrelativistically.

One should investigate the functions $T_f$ and $T_s$ in the energy region
\begin{equation}
I \ll \varepsilon_2 \ll m,
\end{equation}
where both equations describe
the energy distribution, i.e. in the region where corrections to both distributions are expexted to be
small. The
actual analysis show that such region exists. In Fig.5 we show the functions $T_{f,s}$ for characteristic
value $Z=20$. The two descriptions overlap for the energies $\varepsilon_2/m$ between 0.1 and 0.2, i.e.
for the energies $\varepsilon_2 \sim m\alpha Z$ ($\alpha Z=0.146$ for $Z=20$).

Taking into account the Coulomb corrections to the wave function of the outgoing electron in the distribution
$T_f$ one can expand the consistency of the two descriptions to lower energy values. As we saw in previous
Section, the lowest order Coulomb corrections can be taken into account by the factor $g$ given by Eq.(50), i.e by
changing $T_f(x)$ to
\begin{equation}
\tilde{T_f}(\varepsilon_2)=T_f(x)\cdot g(\epsilon).
\end{equation}
The function $\tilde{T_f}(\varepsilon_2)$ is also shown in Fig.~5.

\section{Total cross section}

Now we calculate the total cross section. Following the analysis of the previous Section we present
\begin{equation}
\sigma=\sigma_s+\sigma_f,
\end{equation}
with the two terms on the RHS corresponding to slow and fast ionized electrons
\begin{equation}
\sigma_s=\frac{\alpha r_e^2}{I}\int_0^{\varepsilon_0}d\varepsilon_2T_s(\frac{\varepsilon_2}{I}); \quad
\sigma_f=\frac{\alpha r_e^2}{m}\int_{\varepsilon_0}^{\omega}
d\varepsilon_2T_f(\frac{\varepsilon_2}{m}),
\end{equation}
with $\varepsilon_0$ belonging to the interval determined by Eq.(52).
Since $T_f(x)$ drops as $\ln x/x^2$ at $x \rightarrow \infty $-(Eq.(19)),
the contribution  $\sigma_f$ has a finite value at $\omega \rightarrow \infty$.
Using Eqs.(18) and (42) we find
\begin{equation}
\sigma_s=\frac{14}{9}\cdot \alpha r_e^2(\ln\frac{\varepsilon_0}{I}+c_s); \quad
\sigma_f=\frac{14}{9}\cdot \alpha r_e^2(\ln\frac{m}{\varepsilon_0}+c_f).
\end{equation}
%The logarithmic terms are determined by $I \ll \varepsilon_2 < \varepsilon_0$ and $\varepsilon_0 <\varepsilon_2 \ll m$,
The contributions $c_s$ and $c_f$ come from the regions $\varepsilon_2 \sim I$ and $\varepsilon_2 \sim m$
correspondingly. Thus the total cross section can be written as
\begin{equation}
\sigma=\frac{14}{9}\cdot \alpha r_e^2(\ln\frac{m}{I}+C),
\end{equation}
with $C=c_s+c_f$. In this approach $C$ can exhibit a weak dependence on $Z$.

Note that the low energy contribution to $c_s$ can be calculated as
\begin{equation}
c_s=  \int_0^ \infty d\epsilon(T_s(\epsilon)-\tilde{T_s}(\epsilon)),
\end{equation}
with the functions $T_s$ and $\tilde{T_s}$ being determined by Eqs.(39) and (51). The integral is saturated
by $\epsilon \lesssim 1$, providing the value $c_s=-0.027$.

The presentation (57) for the cross section can be obtained by noting that the function (53) approximates
the energy distribution well enough. The largest deviations from the exact curve are of the order of several
percent, taking place at $\varepsilon_2 \lesssim I$. Using Eq.(20) we find that
$\tilde{T_f}(\varepsilon_2)= \frac{14}{9}\frac{m}{\varepsilon_2+I}$ at $\varepsilon_2 \ll m$,
providing a logarithmic term on the RHS of Eq.(57). Since at $\varepsilon_2 \sim I$ we can put  $\tilde{T_f}(\varepsilon_2)
=\frac{m}{I}\tilde{T_s}(\epsilon)$, the parameter $c_s$ is determined by Eq.(58).
The actual numerical calculations employing the function $\tilde T_f(\varepsilon_2)$
provide values of $C$ changing from
1.23 for $Z=1$ to 1.31 for $Z=50$. Note that the integral over large energies converges slowly
due to a rather slow drop of the function $T_f(x)$ (Eq.(19)). For the
characteristic value $Z=20$ we obtain $c_f=C-c_s=1.27$,
putting $\omega=\infty$ as the upper limit of the second integral on the RHS of Eq.(55). However assuming the
upper limits of integration to be $5m$ or $10m$ we find values of $c_f$ to be 0.67 and 0.91 correspondingly.

For the ground state of a not very heavy (Eq.(2)) single-electron ion we have
$I=m(\alpha Z)^2/2$ and Eq.(57) can be written as
\begin{equation}
\sigma=\frac{14}{9}\cdot \alpha r_e^2(\ln\frac{2}{(\alpha Z)^2}+C).
\end{equation}

If the parameter $(\alpha Z)^2$ is not treated as a small one, one should use relativistic Coulomb functions for
all the electrons and positron. It is known \cite{BM} that the ultrarelativistic particles of the
$e^-e^+$ pair can be
described by Furry-Sommerfeld-Maue  (FSM) functions \cite{FS} which provide a relative accuracy
 $(\alpha Z)^2/\ell$
with $\ell$ standing for the orbital momenta. Since the pair transfers momentum $q \sim m$ to the nucleus or
to the outgoing electron the values of $\ell  \sim p_i/q \sim \omega/m$ are important. Thus corrections to FSM functions
can be neglected. The calculation for the pair creation in the field of the nucleus \cite{BM} resulted in an additional
contribution $f(Z)$ to the cross section, which does not depend on the photon energy,
and can be presented as $(\alpha Z)^2$ series. In our process the ultrarelativistic particles of the pair can be considered
in similar way, providing the same contribution $f(Z)$ to the cross section.

However the bound electron and the ionized electron at $\varepsilon_2 \sim m$
should be described by totally relativistic Coulomb functions at $\alpha Z \sim 1$. One can employ the presentation
\cite{GR} in which relativistic functions are expressed in terms of $(\alpha Z)^2$ series with the FSM functions as
zero order terms. Hence in the case $\alpha Z \sim 1$ the cross section can be written as
\begin{equation}
\sigma=\frac{14}{9}\cdot \alpha r_e^2(\ln\frac{m}{I}+C_R),
\end{equation}
with $C_R=C+(\alpha Z)^2\delta C$, while $\delta C$ can be presented as a certain $(\alpha Z)^2$ series.

Until now we considered a single electron ion.
Turning to the case of an atom containing $Z$ electrons or of an ion containing $n_e$ electrons one
can see that for any bound electron the structure of the cross section is similar to that
for the single-electron case. The contribution $c_f$ caused by large
energies $\varepsilon_2 \sim m$ is the same for all the electrons. We saw the contribution $c_s$
to be numerically small for K electrons. Since the other electrons are less bound, it is still smaller for
them and can be neglected. Thus we can write for a bound electron state with the binding energy $I_b$,
\begin{equation}
\sigma=\frac{14}{9}n_b\cdot \alpha r_e^2(\ln\frac{m}{I_b}+C),
\end{equation}
with $C \approx 1.3$, $n_b$ is the number of electrons in the state $b$.

Note that we omitted all contributions, which drop with $\omega$. The largest neglected terms are of
the order $(\frac{m^2}{\omega \eta_b})^2$, which are much smaller then unity for $\omega \gg m\frac{m}{\eta_b
}$.
This condition is stronger then (1). However, in the case $\omega \sim m^2/\eta_b$ we can still use Eqs.(59)-(61)
in logarithmic approximation due to the large value of $\ln(m^2/\eta_b^2)$.

Note that our cross section reaches a constant value at $\omega \rightarrow \infty$. On the other hand
the cross section of pair creation in the field of the nucleus \cite{2} and
that on a free electron \cite{7,8} increase as $\ln\omega$ in this limit. This happens because the logarithmic terms
are caused by integration over momentum transferred $q$ in the former case and over the momentum of outgoing
electron $p_2$ in the latter case. In both cases the lower limits of integration are of the order
$m^2/\omega$, causing the terms $\ln \omega$ in the cross sections. In our case the effects of the binding are
important at the lower limits and we obtain $\ln 1/(\alpha Z)^2$ instead.

In \cite{data} ionization of internal shells of silver and gold atoms in coincidence with pair creation was measured
for 1 GeV photons. Using Eq.(61) we find the cross sections for ionization of the K shells
to be $7.8~mb$ in Ag and $5.9~mb$ in Au. In the
latter case $Z=79$ and the errors can be about $30\%$. The experimental results are $18 \pm 6~mb$ for Ag and
$8.3 \pm 6.2 ~mb$ for Au. Our result for ionization of the L-shell in Au is $37~mb$, while the experiment provides
$116 \pm 76~ mb$.
Thus our calculations underestimate the experimental data for silver, being in agreement
with the results for gold.

Now we find the values of the photon energy $\omega$ for which the considered process is the main mechanism of
formation ions. We must compare the cross section of our process to the asymptotics of photoionization
and to that of the  Compton scattering on the bound electrons. For small and moderate values of $Z$ the cross section
of the latter process is larger than that of the former one at $\omega \gg m$. The total cross section
of the high energy Compton scattering on a bound electron is equal to that on a free electron \cite{MG}. The asymptotics
of the latter is
$$ \sigma_C(\omega)=\pi r_e^2\frac{\ln(2y)+1/2}{y},$$
with $y=\omega/m$. Thus $\sigma_C$ drops while the photon energy increases. At certain value $\omega_0$
\begin{equation}
\sigma_C(\omega_0)=\sigma,
\end{equation}
and $\sigma>\sigma_C(\omega)$ for $\omega>\omega_0$.
In Fig 6. we show the $Z$ dependence of $\omega_0$ for the K electrons of single electron ions and of atoms.
The value for hydrogen $\omega_0= 73.6 MeV$ is the smallest one.
For the
external electrons the values of $\omega_0$ become still smaller due to the small values of the binding energies.
For example the binding energies of $3s$ and $4s$ electrons in Na and K are $4.9 eV$ and $4.1 eV$ correspondingly
 \cite{FK}, providing the values $\omega_0=65.6 MeV$ and $\omega_0=66.7 MeV$.

\section{Summary}

We analysed formation of ions by high energy photons accompanied by creation of $e^-e^+$ pairs.
We calculated the energy distributions $d\sigma /d\varepsilon$ for creation of an
ion and a continuum electron with kinetic
energy $\varepsilon \ll \omega$. We showed that the  slow electrons with $\varepsilon$ being of the order
of the binding energies $I_b$ and the fast electrons with the energies $\varepsilon \gg I$ need a separate treatment.
We carried out matching of the two regions and found analytical formula (53), which approximates the whole spectrum
of the outgoing electrons.

We integrated the energy distributions and found the expressions (59) and (61) describing the cross sections for the ionization
of single-electron ions and of any state in a many-electron atom. We showed that the high energy asymptotics of the cross
sections does not depend on the photon energy. We found the values of the photon energies
$\omega_0$ for which ionization accompanied
by pair creation becomes to be the dominant mechanism for the formation of an ion. The $Z$-dependence of $\omega_0$
for K electrons is shown in Fig.~6. The value of $\omega_0$ appeared to be about $74 MeV$ in hydrogen, increasing
with $Z$, and being somewhat smaller for loosely bound external electrons of heavier atoms.

We carried out calculations for not very heavy atoms-see Eq.(2). The approach can be generalized for the case
$\alpha Z \sim 1$ as well.

Note that related problem of the influence of atomic electrons on the pair creation was considered in \cite{W},
\cite{mx}. The authors focused on  modification of characteristics of the created pair by the atomic field.
That is why they used some additional approximations in the description of atomic electrons. However, the totally
integrated cross section which includes all inelastic transitions, presented in \cite{o} for the case of hydrogen ($\sigma=\alpha r_e^2\cdot 19$),
can be compared with our result $\sigma=\alpha r_e^2\cdot 18$.

As far as we know the only related experiment was carried out in \cite{data}. Our results are consistent with these data.
There are still large errors in experimental and theoretical analysis. This should stimulate further development of both.

\subsection*{Acknowledgments}

The work was partially supported by the DFG grant 436 RUS 113/822/0-1. The work of Kh.R. was supported
by INTAS YS grant 04-83-2674 and by Uzbek Academy of Sciences grant 66-06. Four of us (E. G. D., A. I. M., I. A. M. and
Kh. R.) thank for hospitality during
their visits to the Justus-Liebig University of Giessen.

\newpage

\subsection*{Figure captions}

\noindent{\bf Fig.~1.} Feynman diagrams describing creation of $e^-e^+$ pair in the
field of the nucleus by the photon. Wavy line shows the photon, solid lines
show electron and positron, dashed line stands for the interactions between the
created pair and the nucleus.

\noindent{\bf Fig.~2.} Feynman diagrams describing creation of $e^-e^+$ pair accompanied
by removing of the bound electron (denoted by a dark blob) to continuum state with asymptotic
momentum $p_2$. Other notations are the same as in Fig.1.

\noindent{\bf Fig.~3.} The function $T_f(x)$ describing energy distributions of
fast electrons as defined by Eq.(19), with $x$ standing for the electron kinetic
energy in units of the electron mass.

\noindent{\bf Fig.~4.} a) The function $T_s(\epsilon)$ describing energy distributions of
slow electrons as defined by Eq.(39), with $\epsilon$ standing for the electron kinetic
energy in units of the ionization potential; b)The energy dependence of the
difference $\delta T_s= \tilde{T_s}-T_s$
between approximate function $\tilde{T}_s$ defined by Eq.(51) and the function $T_s$.

\noindent{\bf Fig.~5.} Matching of the two regions of the electron spectrum for the characteristic case
$Z=20$. a) Dashed line shows the function $T_f(x)$ defined by Eq.(19), calculated for the fast electrons
with $x=\varepsilon_2/m$.
Dotted line shows the function $\frac{m}{I}T_s(\frac{mx}{I})$, describing distribution of slow electrons. Solid line
shows the function $\tilde{T_f}(x)$ defined by Eq.(53). b) Here we show the lower part of the spectrum in more
detail (the energy value $\varepsilon_2=m(\alpha Z)^2/2=5.4 keV$ corresponds to $x \approx 1.07 \cdot 10^{-2}$).

\noindent{\bf Fig.~6.} Dependence of the photon energy $\omega_0$ on the value of the nuclear charge $Z$.
At $\omega>\omega_0$ the ionization accompanied by creation of $e^-e^+$ pairs is the dominant mechanism
of the K shell ionization. Curve 1 is for the single-electron ions, curve 2 is for atoms with Z electrons.

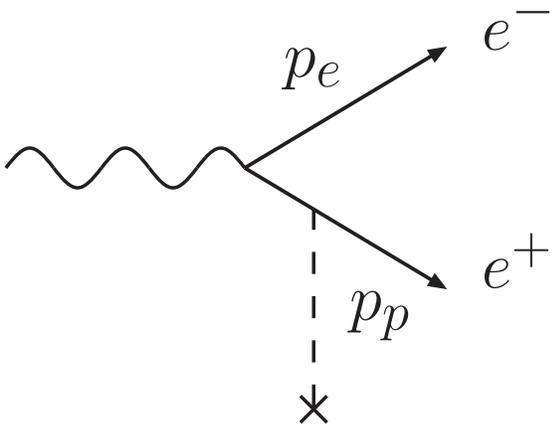
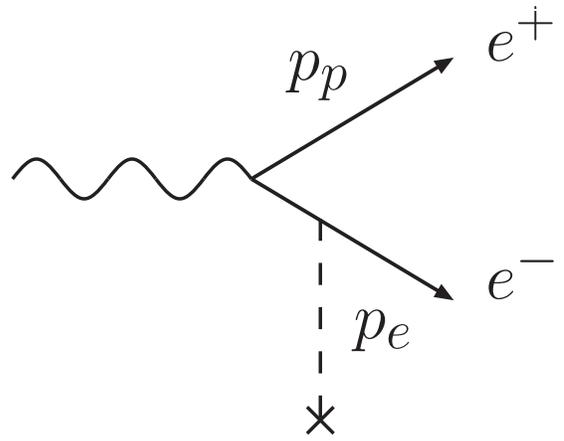
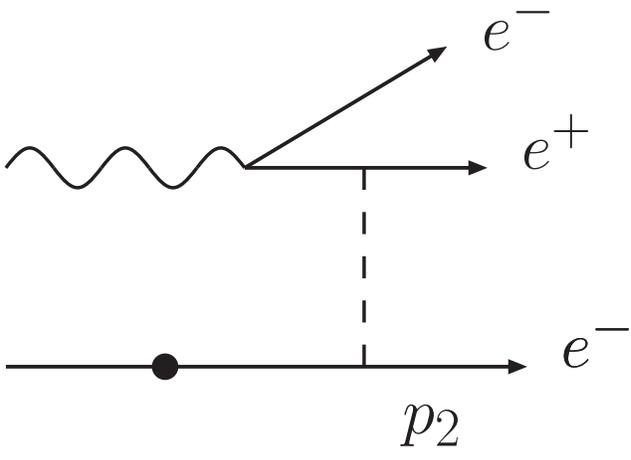
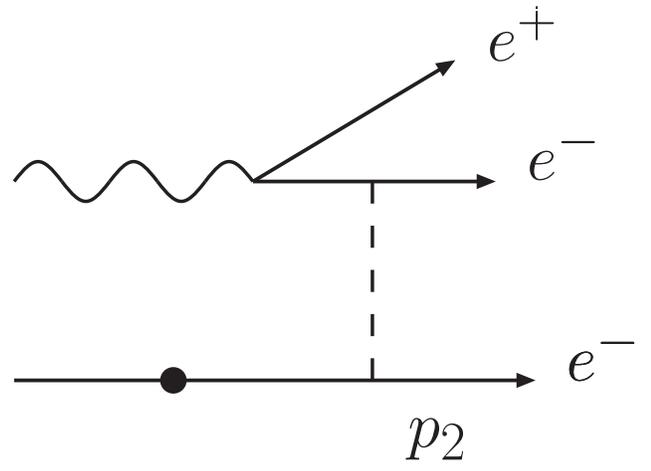
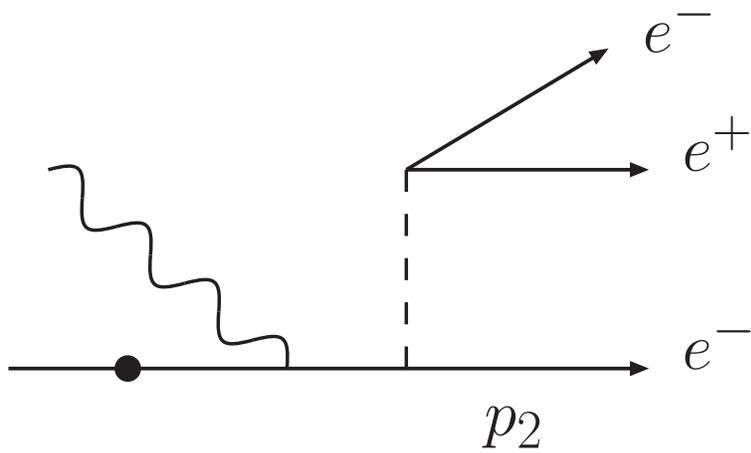
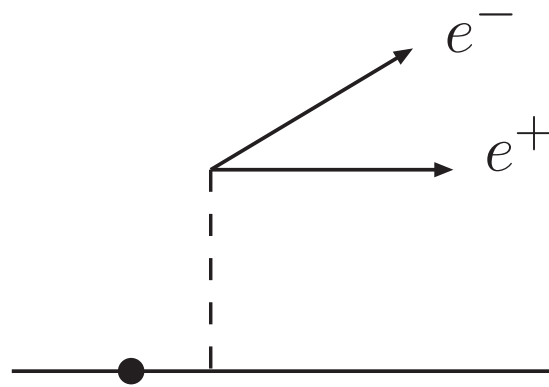
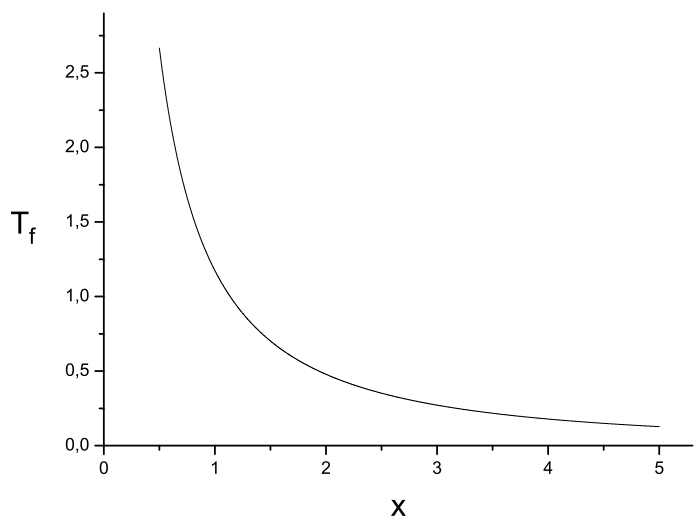
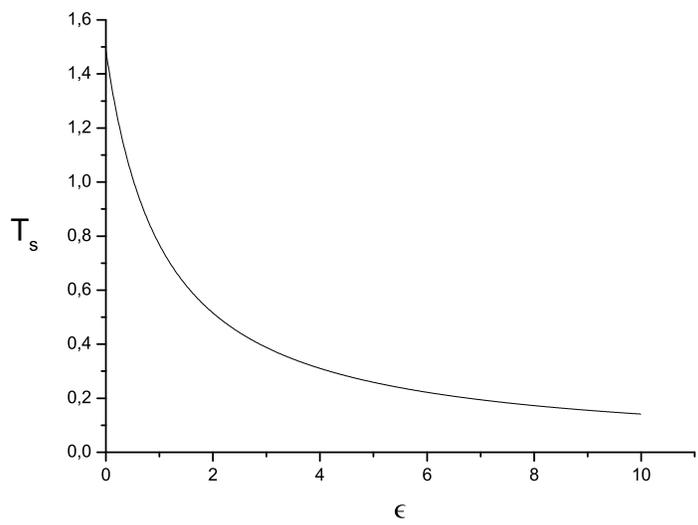
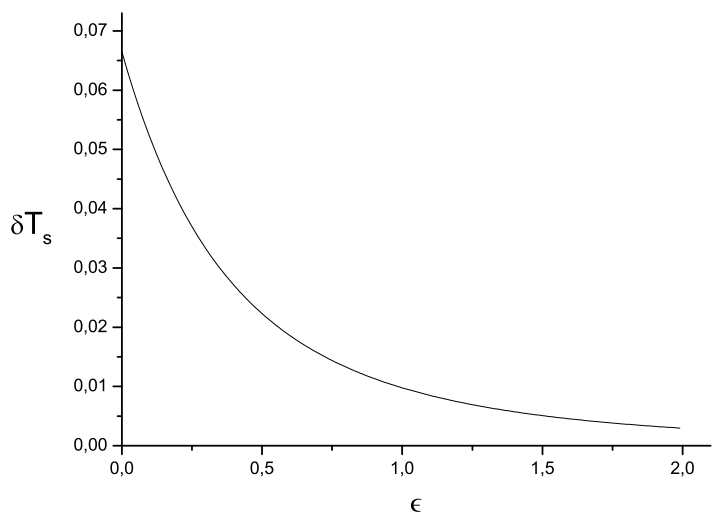
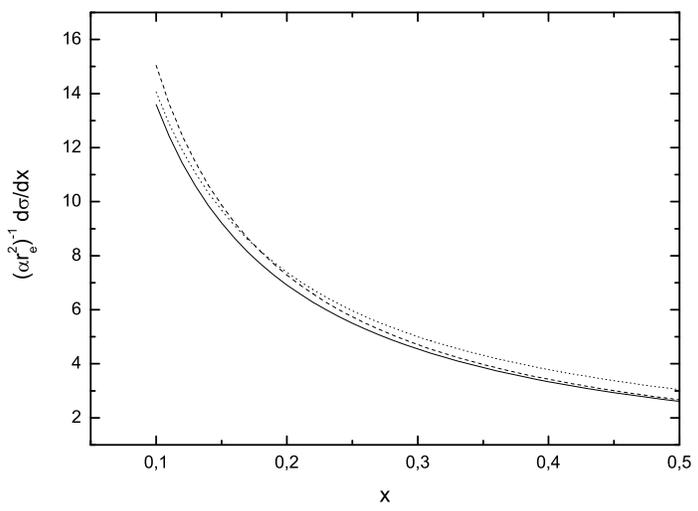
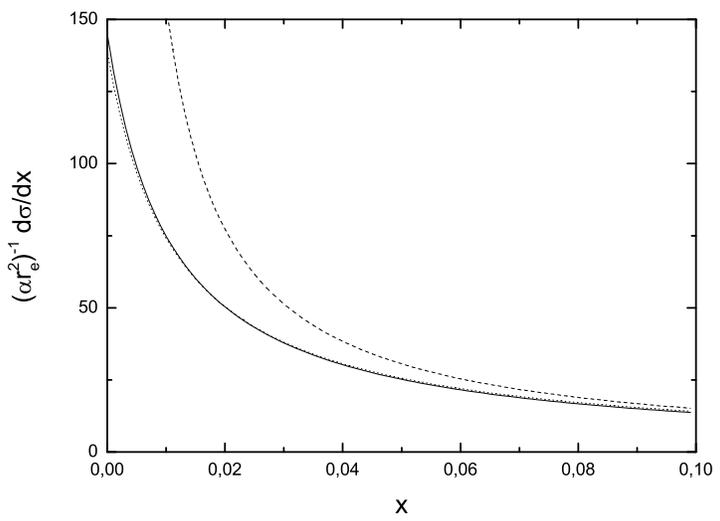
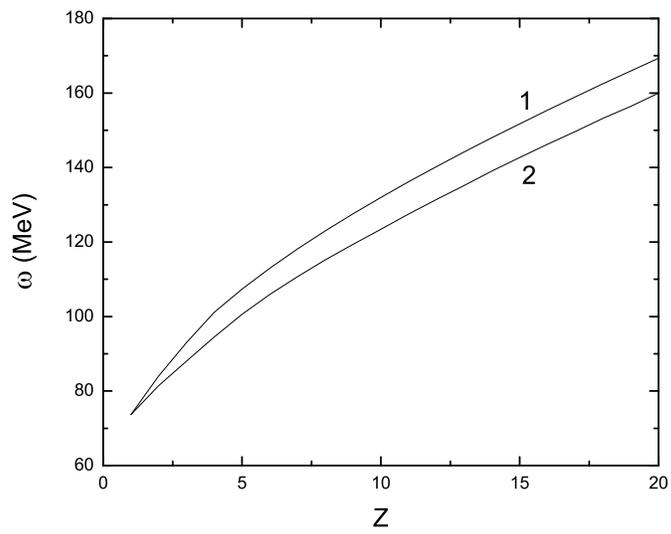
\end{document}